\documentclass[review]{elsarticle}

\usepackage{hyperref} 
\usepackage{graphicx}
\usepackage{bm}
\usepackage{amsmath,amssymb,amsfonts}
\usepackage{amsthm}
\usepackage{derivative}
\usepackage{color}
\usepackage{balance}

\usepackage[margin=1.0in]{geometry}

\newtheorem{remark}{Remark}
\newtheorem*{theorem*}{Theorem}
\newtheorem*{definition*}{Definition}

\usepackage{fancyhdr}

\journal{Journal of \LaTeX\ Templates}

\begin{document}

\begin{frontmatter}

\title{Off-Grid Direction-of-Arrival Estimation Using Second-Order Taylor Approximation}
\tnotetext[mytitlenote]{The work of H. Huang is supported by the Graduate School CE within the Centre for Computational Engineering at Technische Universität Darmstadt.}


\author[mymainaddress]{Huiping Huang} 
\ead{h.huang@spg.tu-darmstadt.de}

\author[mysecondaryaddress]{Hing Cheung So}
\ead{hcso@ee.cityu.edu.hk}

\author[mymainaddress]{Abdelhak M. Zoubir}
\ead{zoubir@spg.tu-darmstadt.de}

\address[mymainaddress]{Department of Electrical Engineering and Information Technology, Technische Universität Darmstadt, Darmstadt, Germany}
\address[mysecondaryaddress]{Department of Electrical Engineering, City University of Hong Kong, Hong Kong, China}

\begin{abstract}
The problem of off-grid direction-of-arrival (DOA) estimation is investigated. We develop a grid-based method to jointly estimate the closest \textit{spatial frequency} (the sine of DOA) grids, and the gaps between the estimated grids and the corresponding frequencies. By using a second-order Taylor approximation, the data model under the framework of joint-sparse representation is formulated. We point out an important property of the signals of interest in the model, namely the proportionality relationship, which is empirically demonstrated to be useful in the sense that it increases the probability of the mixing matrix satisfying the block restricted isometry property. Simulation examples demonstrate the effectiveness and superiority of the proposed method against several state-of-the-art grid-based approaches.
\end{abstract}


\end{frontmatter}


\section{Introduction}
\label{introduction}
\setcounter{page}{1}

Grid-based methods have gained interest in direction-of-arrival (DOA) estimation in recent years. Such approaches include least absolute shrinkage and selection operator (LASSO) \cite{Tibshirani1996, Malioutov2005, Weiss2014} and sparse iterative covariance-based estimation \cite{Stoica2011a, Stoica2011b, Stoica2012, Babu2014}, among others. See \cite{Yang2018} for a comprehensive review of grid-based sparse methods for DOA estimation. The advantage of grid-based methods is that they have super-high resolution even in the case when only one single snapshot is available, provided that all the source spatial frequencies align exactly with the preset grid. However, this condition may not be satisfied in practice, since the region of interest (ROI) contains infinite candidates and hence grid mismatch almost always exists when we split the ROI into a finite number of grids. This is known as the off-grid issue and has attracted a lot of research interest in array signal processing during the past decade, see for example \cite{Zhu2011, Yang2012, Yang2013, Duarte2013, Jagannath2013, Fortunati2014, Tan2014, Dai2017, Walewski2017, Liu2017, Abtahi2018, Wang2018, Wu2018, Zhou2018, Wang2019, Wan2021, Ma2021}.

Existing solutions to tackle the off-grid problem can be categorized into three groups. The first group uses denser grids or the coarse-to-fine strategy such as \cite{Malioutov2005}. The drawbacks of these methods are twofold. On one hand, denser grids lead to extremely expensive computational complexity; on the other hand, too dense grids may result in weak incoherence among the steering vectors. The second group consists of the so-called \textit{gridless} approach \cite{Tang2013, Yang2015, Steinwandt2016, Steffens2017, Wagner2019, Zhang2021}. Its weakness is that most of these methods are restricted to regularly sampled measurements that can only be taken from a uniform linear array (ULA) \cite{Wagner2021}. The last group of methods estimates the off-grid bias together with the grids closest to the true spatial frequencies. Representative works include the first-order Taylor approximation \cite{Jagannath2013, Tan2014} and the neighbor-grid based method \cite{Abtahi2018}, denoted in this paper as 1st Taylor G-LASSO and Neighbor G-LASSO, respectively.

It is known that in general the first-order Taylor approximation is accurate enough, especially when the grid size is small. However, when the grid size is set not small enough so as to save computational cost, there still exists a large bias. In such a situation, a high-order Taylor approximation decreases the approximation error. To this end, we introduce a second-order Taylor approximation in off-grid DOA estimation. We observe in this case the proportionality relationship of the signals of interest. With this, we propose a novel optimization approach which is shown by simulation to produce more accurate frequency estimates in off-grid scenarios. {Moreover, the uniqueness issue of the proposed method is discussed by means of the restricted isometry property (RIP), which is one of the most important tools in compressive sensing \cite{Eldar2009}.}

{\textit{Notation:} In this paper, bold-faced lower-case and upper-case letters stand for vectors and matrices, respectively. Superscripts $\cdot^{\text{T}}$, $\cdot^{\text{H}}$, and $\cdot^{*}$ denote transpose, Hermitian transpose, and complex conjugate operators, respectively. $\text{vec}\{ \cdot \}$ denotes the vectorization operator, $\text{diag}\{\cdot\}$ returns a diagonal matrix whose main diagonal is given in the curly bracket, and $\rm{Re}\{\cdot\}$ and $\rm{Im}\{\cdot\}$ are real and imaginary parts of a complex-valued variable, respectively. $\odot$ symbolizes the Khatri-Rao product. $\mathbb{C}$ and $\mathbb{R}$ are the sets of complex and real numbers, respectively. ${\bf I}$ is the identity matrix of appropriate dimension. ${\bf 0}$ and ${\bf 1}$ denote the all-zeros and the all-ones vectors of appropriate length, respectively. For a vector ${\bf x}$, $|{\bf x}|$ and $\|{\bf x}\|_{2}$ represent the element-wise absolute value and the $L_{2}$ norm of ${\bf x}$, respectively. The symbols $\preceq$, $\succeq$, and $\succ$ are element-wise less than or equal to, greater than or equal to, and greater than operators, respectively.  }

\section{Signal Model}
\label{signalmodel}

Suppose a linear array of $M$ sensors whose positions are contained in ${\bf q} = [q_{1}, q_{2}, \cdots, q_{M}]^{\text{T}}$, receives $K$ far-field narrowband signals from directions ${\bm \phi} = [\phi_{1}, \phi_{2}, \cdots, \phi_{K}]^{\text{T}}$ with $\phi_{k} \in [-\pi/2 , \pi/2)$. For simplicity, we define the spatial frequencies as ${\bm u} = [u_{1}, u_{2}, \cdots, u_{K}]^{\text{T}}$ with $u_{k} = \sin(\phi_{k}) \in [-1 , 1)$. The array observation can be modeled as 
\begin{align*}
	{\bf y} = \sum_{k = 1}^{K}s_{k}{\bf a}(u_{k}) + {\bf n} = {\bf A}({\bm u}){\bf s} + {\bf n},
\end{align*}
where $s_{k}$ is the $k$-th signal waveform, ${\bf s} = [s_{1}, s_{2}, \cdots, s_{K}]^{\text{T}}$ represents the signal vector, and ${\bf n} \in \mathbb{C}^{M}$ is the noise vector. The steering matrix ${\bf A}({\bm u}) = [{\bf a}(u_{1}), {\bf a}(u_{2}), \cdots, {\bf a}(u_{K})] \in \mathbb{C}^{M \times K}$ has the steering vectors as columns, where ${\bf a}(u_{k}) = [e^{\jmath \frac{2\pi q_1}{\lambda}u_{k}}, e^{\jmath \frac{2\pi q_{2}}{\lambda}u_{k}}, \cdots, e^{\jmath \frac{2\pi q_{M}}{\lambda}u_{k}}]^{\text{T}}$,
for $k = 1, 2, \cdots, K$, with $\lambda$ being the signal wavelength and $\jmath = \sqrt{-1}$. 

In grid-based methods, we formulate the signal model by means of a sparse representation, as
\begin{align*}
	{\bf y} = \sum_{l=1}^{L} x_{l}{\bf a}(v_{l}) + {\bf n} = {\bf A}({\bm v}){\bf x} + {\bf n},
\end{align*}
where ${\bm v} = [v_{1}, v_{2}, \cdots, v_{L}]^{\text{T}}$ denotes the frequency grid vector with $L$ being the number of grids (in general $L \gg M > K$), ${\bf A}({\bm v}) \in \mathbb{C}^{M \times L}$ stands for the overcomplete dictionary matrix, and ${\bf x} = [x_{1}, x_{2}, \cdots, x_{L}]^{\text{T}}$ is a sparse vector whose elements $x_{l} = s_{k}$ if $v_{l} = u_{k}$, and $x_{l} = 0$ otherwise. When the true frequencies do not exactly lie in the preset grids, we encounter the off-grid issue. To handle this problem, we propose a method to simultaneously estimate the closest frequency grids, and the gaps between the closest grids and the true frequencies, using a second-order Taylor approximation.

\section{Proposed Method}
\label{proposedmethods}

\subsection{Second-Order Taylor Approximation}
\label{2ndTaybor}

We start by considering a second-order Taylor approximation of the steering vectors. For any $u_{l}$, we have
\begin{align*}
	{\bf a}({u_{l}}) \approx {\bf a}(v_{l}) + {\bf a}'(v_{l})p_{l} + \frac{{\bf a}''(v_{l})}{2}p_{l}^{2},
\end{align*}
where $v_{l}$ is the grid closest to $u_{l}$, ${\bf a}'(v_{l}) = \left. \odv{{\bf a}(v)}{v} \right |_{v = v_{l}} $, ${\bf a}''(v_{l}) = \left. \odv[2]{{\bf a}(v)}{v} \right |_{v = v_{l}}$, and $p_{l} = u_{l} - v_{l} \in [-\delta/2 , \delta/2]$ with $\delta$ being the grid size. Collecting all the candidates, we have
\begin{align*}
	[{\bf a}(u_{1}), \! \cdots \! , {\bf a}(u_{L})] \approx {\bf A}({\bm v}) \!+\! {\bf A}'({\bm v})\text{diag}\{{\bf p}\} \!+\! \frac{1}{2}{\bf A}''({\bm v})\text{diag}\{{\bf p}\}^{2},
\end{align*}
where ${\bf A}'({\bm v}) = [{\bf a}'(v_{1}), \cdots, {\bf a}'(v_{L})] \in \mathbb{C}^{M \times L}$, ${\bf A}''({\bm v}) = [{\bf a}''(v_{1}), \cdots, {\bf a}''(v_{L}) ] \in \mathbb{C}^{M \times L}$, and ${\bf p} = [p_{1}, p_{2}, \cdots, p_{L}]^{\text{T}}$. Hence, the signal model can be approximately written as:
\begin{align}
\label{signalmodel_2nd}
	{\bf y} \approx & \left[ {\bf A}({\bm v}) + {\bf A}'({\bm v})\text{diag}\{{\bf p}\} + \frac{1}{2}{\bf A}''({\bm v})\text{diag}\{{\bf p}\}^{2} \right] {\bf x} + {\bf n} \nonumber \\
		= & \left[ {\bf A}({\bm v}) , {\bf A}'({\bm v}) , \frac{1}{2}{\bf A}''({\bm v}) \right] \!\! \left[ \!\!
			  \begin{array}{c}
			  {\bf x} \\
			  \text{diag}\{{\bf p}\}{\bf x} \\
			  \text{diag}\{{\bf p}\}^{2}{\bf x}
			  \end{array} \!\! \right] + {\bf n},
\end{align}
where the signals of interest $[{\bf x}^{\text{T}}, ( \text{diag}\{{\bf p}\}{\bf x})^{\text{T}}, (\text{diag}\{{\bf p}\}^{2}{\bf x})^{\text{T}}]^{\text{T}}$ are referred to as block signal in the sequel. \vspace{-0.4mm}

\subsection{Properties of the Block Signal}
\label{property}

As shown in signal model (\ref{signalmodel_2nd}), the unknown block signal is divided into three parts: (i) ${\bf x}_{1} \triangleq {\bf x}$, (ii) ${\bf x}_{2} \triangleq \text{diag}\{{\bf p}\}{\bf x}$, and (iii) ${\bf x}_{3} \triangleq \text{diag}\{{\bf p}\}^{2}{\bf x}$. {Without loss of generality, we assume ${\bf x}$ is a real-valued vector, see Remark \ref{imaginary_to_real} below.} Denote the $l$-th entries of ${\bf x}_{1}$, ${\bf x}_{2}$, and ${\bf x}_{3}$ as $x_{1,l}$, $x_{2,l}$, and $x_{3,l}$, respectively. We notice the following properties of the block signal $[{\bf x}_{1}^{\text{T}}, {\bf x}_{2}^{\text{T}}, {\bf x}_{3}^{\text{T}}]^{\text{T}}$.
\begin{itemize}
	\item Since ${\bf x}$ is a sparse vector as mentioned in Section \ref{signalmodel}, ${\bf x}_{1}$, ${\bf x}_{2}$, and ${\bf x}_{3}$ are all sparse and share the same sparsity pattern. This property is known as block-sparsity \cite{Jagannath2013} or joint-sparsity \cite{Tan2014}. 
	
	\item It holds that $x_{2,l} = p_{l}x_{1,l}$ and $x_{3,l} = p_{l}^{2}x_{1,l}$. Due to $-\delta/2 \leq p_{l} \leq \delta/2, ~ \forall l \in \{1, 2, \cdots, L\}$, it is easy to verify that the following inequalities hold:
	\begin{align}
		\label{x123}
		-\frac{\delta}{2}\left|{\bf x}_{1}\right| \preceq {\bf x}_{2} \preceq \frac{\delta}{2}\left|{\bf x}_{1}\right|, ~~
		-\left(\frac{\delta}{2}\right)^{2}\left|{\bf x}_{1}\right| \preceq {\bf x}_{3} \preceq \left(\frac{\delta}{2}\right)^{2}\left|{\bf x}_{1}\right|. 
	\end{align}
	
	\item It can be seen that ${\bf x}_{1}$, ${\bf x}_{2}$, and ${\bf x}_{3}$ satisfy the proportionality relationship, as
	\begin{align}
	\label{proportional}
		x_{2,l}^{2} = x_{1,l}x_{3,l}, ~ \forall l \in \{1, 2, \cdots, L\}.
	\end{align}
	\end{itemize}

{
\begin{remark}
\label{imaginary_to_real}
For any complex-valued data model, say ${\bf y} = {\bf A}{\bf x}$, we have its real-valued counterpart as ${\bf{\tilde y}} = {\bf{\tilde A}}{\bf{\tilde x}}$, where ${\bf{\tilde y}} = [\rm{Re}\{{\bf y}\}^{\rm{T}} , \rm{Im}\{{\bf y}\}^{\rm{T}}]^{\rm{T}}$, ${\bf{\tilde x}} = [\rm{Re}\{{\bf x}\}^{\rm{T}} , \rm{Im}\{{\bf x}\}^{\rm{T}}]^{\rm{T}}$, and ${\bf{\tilde A}} = \left[ \begin{array}{cc}
\rm{Re}\{{\bf A}\} & -\rm{Im}\{{\bf A}\} \\
\rm{Im}\{{\bf A}\} & \rm{Re}\{{\bf A}\}
\end{array} \right]$.
\end{remark}
}

\subsection{Problem Formulation Development}
\label{proposedminiprob}

Based on the aforementioned relationships among ${\bf x}_{1}$, ${\bf x}_{2}$, and ${\bf x}_{3}$, we propose the following minimization problem:
\begin{align}
\label{proposednonconvex}
	\min_{{\bf x}_{1}, {\bf x}_{2}, {\bf x}_{3}} ~~ g({\bf x}_{1}, {\bf x}_{2}, {\bf x}_{3})  \qquad \text{s.t.} ~~ \text{(\ref{x123})} ~ \text{and} ~ \text{(\ref{proportional})}.
\end{align}
The cost function in (\ref{proposednonconvex}) is given by
\begin{align}
	g({\bf x}_{1}, {\bf x}_{2}, {\bf x}_{3}) \triangleq \frac{1}{2}\left\| {\bf y} \!-\! {\bf A}({\bm v}){\bf x}_{1} \!-\! {\bf A}'({\bm v}){\bf x}_{2} \!-\! \frac{1}{2}{\bf A}''({\bm v}){\bf x}_{3} \right\|_{2}^{2} + \mu \left\| \left[ {\bf x}_{1}^{\text{T}} , {\bf x}_{2}^{\text{T}}, {\bf x}_{3}^{\text{T}} \right] ^{\text{T}} \right\|_{2,1},
\end{align}
where $\mu$ is a regularization parameter balancing the data fitting and the model sparsity, and $\left\| \cdot \right\|_{2,1}$ is the mixed $L_{2,1}$ norm of a vector, defined as
\begin{align*}
	\left\| \left[ {\bf x}_{1}^{\text{T}} , {\bf x}_{2}^{\text{T}}, {\bf x}_{3}^{\text{T}} \right]^{\text{T}} \right\|_{2,1} = \sum_{l = 1}^{L}\sqrt{|x_{1,l}|^{2} + |x_{2,l}|^{2} + |x_{3,l}|^{2}}.
\end{align*}

{Problem (\ref{proposednonconvex}) is non-convex and hard to solve due to its constraints. We first consider the constraints of (\ref{x123}). The difficulty of dealing with (\ref{x123}) comes from the absolute value operator \cite{Tan2014}. However, when the signals are assumed to be real positive, i.e., ${\bf s} \succ {\bf 0}$ (and ${\bf x}_{1} = {\bf x} \succeq {\bf 0}$), the constraints of (\ref{x123}) in (\ref{proposednonconvex}) become}
\begin{align}
\label{constraint_postive}
	-\frac{\delta}{2}{\bf x}_{1} \preceq {\bf x}_{2} \preceq \frac{\delta}{2}{\bf x}_{1}, ~~ {\bf 0} \preceq {\bf x}_{3} \preceq \left(\frac{\delta}{2}\right)^{2}{\bf x}_{1}, ~~ {\bf x}_{1} \succeq {\bf 0},
\end{align}
{which are linear and thus convex.} It is worth pointing out that ${\bf 0} \preceq {\bf x}_{3}$ in (\ref{constraint_postive}) is the result of $x_{3,l} = p_{l}^{2}x_{1,l}, ~ \forall l \in \{1, 2, \cdots, L \}$ and ${\bf x}_{1} \succeq {\bf 0}$. {Note that the assumption of real positive signals is valid in various situations. For instance, in multiple-snapshot scenarios, the signal vector denotes the signal powers which are naturally positive, see Remark \ref{multiple_to_single}. }

In the sequel, we consider the last constraint in (\ref{proposednonconvex}), viz. (\ref{proportional}). Firstly, we convert (\ref{proportional}) to its equivalent form as in \cite{Park2017}:
\begin{align}
\label{soc_equality}
	\left\| \left[ \!\! \begin{array}{c}
		2x_{2,l} \\
		x_{1,l} - x_{3,l}
		\end{array} \!\! \right] \right\|_{2} {= ~\! } x_{1,l} + x_{3,l}, ~ \forall l \in \{1, 2, \cdots, L\}.
\end{align}
Then, we introduce an additional variable ${\bf z} \in \mathbb{R}^{L}$ with entries $z_{l} $ satisfying 
\begin{align}
\label{z}
	0 \leq z_{l} \leq \eta, ~ \forall l \in \{1, 2, \cdots, L\},
\end{align}
where $\eta$ is a small user-defined parameter, and rewrite (\ref{soc_equality}) as
\begin{align}
\label{soc_inequality}
	\!\!\! \left\| \left[ \!\! \begin{array}{c}
		2x_{2,l} \\
		x_{1,l} - x_{3,l}
		\end{array} \!\! \right] \right\|_{2} \leq x_{1,l} + x_{3,l} + z_{l}, ~ \forall l \in \{1, 2, \cdots, L\},
\end{align}
which belongs to the set of standard second-order cone and hence is convex.

By replacing the constraint (\ref{x123}) with (\ref{constraint_postive}) and replacing (\ref{proportional}) with (\ref{z}) and (\ref{soc_inequality}), we finally relax the non-convex problem (\ref{proposednonconvex}) into a convex one, as
\begin{align}
\label{proposedconvex}
	\min_{{\bf x}_{1}, {\bf x}_{2}, {\bf x}_{3}, {\bf z}} ~~ g({\bf x}_{1}, {\bf x}_{2}, {\bf x}_{3}) \qquad \rm{s.t.} ~~ & \rm{(\ref{constraint_postive})}, ~ \rm{(\ref{z})}, ~ \rm{and} ~ \rm{(\ref{soc_inequality})}.
\end{align}


\begin{remark}
\label{multiple_to_single}
Note that the proposed method is developed for the single-snapshot scenario. However, it can be easily extended to the case of multiple snapshots. To be precise, when multiple snapshots are available, we have the covariance matrix ${\bf R} = {\bf A}{\bf R}_{s}{\bf A}^{{\rm{H}}} + \sigma^{2}{\bf I}$, where $\sigma^{2}$ is the noise power. Note that we assume the signals to be uncorrelated with the noise, and the noise components are independent and identically distributed. Vectoring ${\bf R}$ yields
\begin{align}
\label{vecR}
	{\rm{vec}}\{{\bf R}\} = ({\bf A}^{*} \odot {\bf A}) {\bf r}_{s} + \sigma^{2}{\rm{vec}}\{{\bf I}\},
\end{align}
where ${\bf r}_{s}$ is the main diagonal of ${\bf R}_{s}$, denoting the signal powers. The data model (\ref{vecR}) is similar to the signal model introduced in Section \ref{signalmodel}, and therefore, we can develop our method on the basis of (\ref{vecR}).
\end{remark}


To analyze the computational cost, we formulate Problem (\ref{proposedconvex}) under the framework of standard second-order cone programming (SOCP) \cite{Lobo1998}, as
\begin{align*}
	\min_{{\bf x}_{1}, {\bf x}_{2}, {\bf x}_{3}, {\bf z}, {\bf t}} ~ & \sum_{l=1}^{L}{t_{l}} = {\bf 1}^{\rm{T}}{\bf t}  \nonumber \\
		\rm{s.t.} ~~~~~~\! & \text{(\ref{constraint_postive})}, ~ \text{(\ref{z})}, ~ \text{and} ~ \text{(\ref{soc_inequality})},  \nonumber \\
		 ~~~~~~\! & \!\!\!\! \sqrt{|x_{1,l}|^{2} \!+\! |x_{2,l}|^{2} \!+\! |x_{3,l}|^{2}} \leq t_{l}, ~ \forall l \in \{1, 2, \cdots , L\},  \nonumber \\
		& \!\!\!\!\! \left\| {\bf y} \!-\! {\bf A}({\bm v}){\bf x}_{1} \!-\! {\bf A}'({\bm v}){\bf x}_{2} \!-\! \frac{1}{2}{\bf A}''({\bm v}){\bf x}_{3} \right\|_{2} \leq \epsilon,
\end{align*}
where ${\bf t} = [t_{1}, t_{2}, \cdots, t_{L}]^{\text{T}}$ is an auxiliary variable vector, and $\epsilon$ is a tuning parameter related to $\mu$ in (\ref{proposedconvex}). The computational cost of the above problem with implementation of SOCP is $\mathcal{O} \! \left( 9(M+1)L^{2} + 72L \right)$ per iteration, and the number of iterations is bounded above by  $\mathcal{O} ( \sqrt{L} )$ \cite{Lobo1998}. The proposed second-order Taylor approximation method is referred to as 2nd Taylor G-LASSO. The computational complexity of the 2nd Taylor G-LASSO, as well as those of LASSO \cite{Tibshirani1996, Malioutov2005}, Neighbor G-LASSO \cite{Abtahi2018}, and 1st Taylor G-LASSO \cite{Jagannath2013, Tan2014}, are summarized in Table \ref{table_cost}.

\begin{table}[h]
\centering
	\caption{Computational Cost Using SOCP Implementation} 
		\begin{tabular}{|c|c|c|}
			\hline
			\textbf{Method} & \textbf{Cost per Iteration} & \textbf{No. of Iterations}  \\ \hline
			LASSO & $\mathcal{O} \! \left( (M+1)L^{2} \right)$ & $\mathcal{O} ( 1 )$ \\ \hline
			Neighbor G-LASSO & $\mathcal{O} \! \left( 4(M+1)L^{2} + 12L \right)$ & $\mathcal{O} ( \sqrt{L} )$ \\ \hline
			1st Taylor G-LASSO & $\mathcal{O} \! \left( 4(M+1)L^{2} + 28L \right)$ & $\mathcal{O} ( \sqrt{L} )$ \\ \hline
			2nd Taylor G-LASSO & $\mathcal{O} \! \left( 9(M+1)L^{2} + 72L \right)$ & $\mathcal{O} ( \sqrt{L} )$ \\ \hline
		\end{tabular}
		\label{table_cost}
\end{table}

\section{{Uniqueness Property of the Proposed Solution }}
\label{discussion}

{Note that, for underdetermined linear systems, uniqueness of a sparse solution is one of the fundamental problems in compressive sensing \cite{Foucart2013}. In this section, we discuss this issue in view of the proposed signal model in (\ref{signalmodel_2nd}).} To this end, we first introduce the following definition and theorem \cite{Eldar2009}:
\begin{definition*}
An $M \times bL$ block matrix ${\bf D}$ is said to have the block RIP with parameter $\beta_{K}$, if for every $K$ block-sparse vector ${\bf c}$ of length $bL$, it holds that
	\begin{align*}
		(1 - \beta_{K})\|{\bf c}\|_{2}^{2} \leq \|{\bf D}{\bf c}\|_{2}^{2} \leq (1 + \beta_{K})\|{\bf c}\|_{2}^{2}.
	\end{align*}
\end{definition*}
\begin{theorem*}
Let ${\bf y} = {\bf D}{\bf c}_{0}$ be measurements of a $K$ block-sparse vector ${\bf c}_{0}$. If ${\bf D}$ satisfies the block RIP with parameter $\beta_{2K}  <  1$, then there exists a unique block-sparse vector ${\bf c}$ satisfying ${\bf y} = {\bf D}{\bf c}$; and further, if ${\bf D}$ satisfies the block RIP with $\beta_{2K}  < \sqrt{2} - 1$, then the convex optimization problem: $\min_{{\bf c}} ~\!\! \|{\bf c}\|_{2,1}  ~\, \rm{s.t.} ~ {\bf y} = {\bf D}{\bf c} $, has a unique solution and the solution is equal to ${\bf c}_{0}$.
\end{theorem*}

Define ${\bf c}_{0} = [{\bf x}^{\text{T}} , (\text{diag}\{{\bf p}\}{\bf x})^{\text{T}} , (\text{diag}\{{\bf p}\}^{2}{\bf x})^{\text{T}}]^{\text{T}}$ and ${\bf D} = [{\bf A}({\bm v}) , {\bf A}'({\bm v}) , \frac{1}{2}{\bf A}''({\bm v})]$. In the absence of noise, our proposed model in (\ref{signalmodel_2nd}) can be rewritten as: ${\bf y} = {\bf D}{\bf c}_{0}$. Without loss of generality, we denote ${\bf{\bar D}}$ as the column-normalized matrix structured from ${\bf D}$. {Our task is to check whether or not ${\bf{\bar D}}$ satisfies the block RIP with parameter $\beta_{2K} < 1$ and $\beta_{2K} < \sqrt{2} - 1$.} Note that determining the RIP parameter, i.e., $\beta_{2K}$, of a given matrix is in general an NP-hard problem \cite{Eldar2010, Tillmann2014}. In what follows, we introduce a Monte Carlo test to check the condition of the block RIP of ${\bf{\bar D}}$.

\begin{figure}
\centerline{\includegraphics[width=\columnwidth]{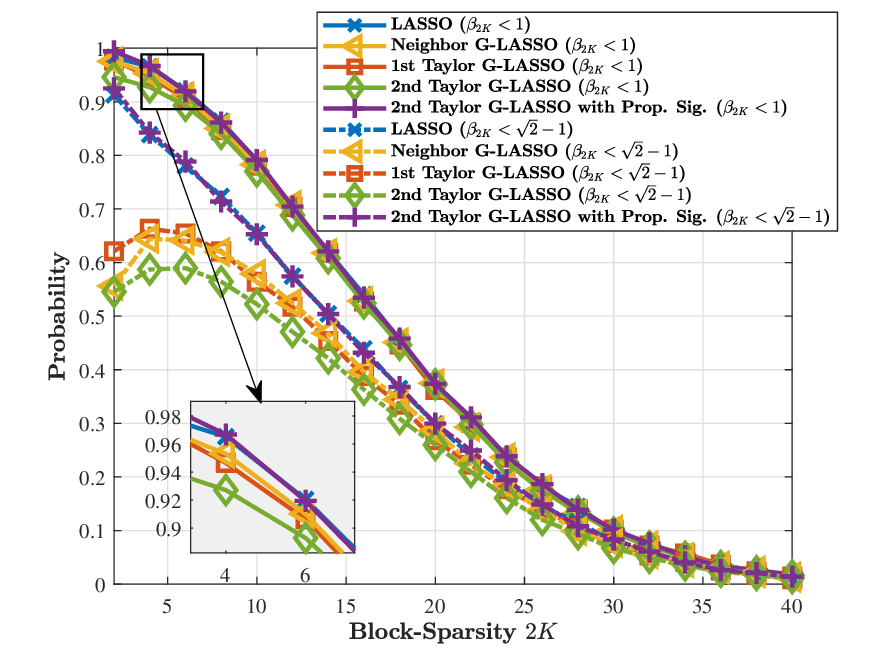}} 
\caption{Empirical probabilities of $\{\beta_{2K} < 1\}$ and $\{\beta_{2K} < \sqrt{2} - 1\}$ versus block-sparsity $2K$ with $10^{4}$ Monte Carlo runs, $M = 8$, and $L = 200$. }
\label{beta_12}
\end{figure}

According to the definition, if ${\bf{\bar D}}$ has the block RIP with parameter $\beta_{2K}$, then for any $2K$ block-sparse vector ${\bf c}$ of length $bL$, it holds that
\begin{align}
\label{RIP_2K}
		(1 - \beta_{2K})\|{\bf c}\|_{2}^{2} \leq \|{\bf{\bar D}}{\bf c}\|_{2}^{2} \leq (1 + \beta_{2K})\|{\bf c}\|_{2}^{2}.
\end{align}
Note that, for any $2K$ block-sparse vector ${\bf c}$, we can write its unit-norm vector as ${\bf{\bar c}} = {\bf c}/\|{\bf c}\|_{2}$, such that $\|{\bf{\bar c}}\|_{2} = 1$. As a result, (\ref{RIP_2K}) becomes:
\begin{align*}
		(1 - \beta_{2K}) \leq \frac{\|{\bf{\bar D}}{\bf c}\|_{2}^{2}}{\|{\bf c}\|_{2}^{2}} = \|{\bf{\bar D}}{\bf{\bar c}}\|_{2}^{2} \leq (1 + \beta_{2K}).
\end{align*}
Based on the above inequalities, the parameter $\beta_{2K}$ is calculated as 
\begin{align}
\label{beta}
	\beta_{2K} = \max \left\{\|{\bf{\bar D}}{\bf{\bar c}}\|_{2}^{2} - 1 \, , 1 - \|{\bf{\bar D}}{\bf{\bar c}}\|_{2}^{2} \right\}.
\end{align}
We randomly generate a unit-norm $2K$ block-sparse vector ${\bf{\bar c}}$, and calculate $\beta_{2K}$ using (\ref{beta}). By repeatedly performing the above steps for $10^{4}$ Monte Carlo runs, we  estimate the empirical probabilities of $\{\beta_{2K} < 1\}$ and $\{\beta_{2K} < \sqrt{2} - 1\}$. The empirical probabilities versus block-sparsity $2K$ are presented in Figure \ref{beta_12}, with $M = 8$, $q_{m} = \frac{(m - 1)\lambda}{2}$ ($m = 1, 2, \cdots, M$), $L = 200$, and $b = 1$ for LASSO, $b = 2$ for Neighbor G-LASSO and 1st Taylor G-LASSO, and $b = 3$ for 2nd Taylor G-LASSO. It is seen that when the block-sparsity is small (less than 8), the probabilities of $\{\beta_{2K} < 1\}$ of all the tested methods are high (greater than 0.9), and their probabilities of $\{\beta_{2K} < \sqrt{2} - 1\}$ are larger than 0.5. Note that in Figure \ref{beta_12}, the plot of 2nd Taylor G-LASSO with proportional signals (abbreviated as ``Prop. Sig.'' in the figure), i.e., (\ref{proportional}), has the highest probability. This reveals that the proportionality relationship of the block signal contains useful information in the sense that it increases the probabilities of $\{\beta_{2K} < 1\}$ and $\{\beta_{2K} < \sqrt{2} - 1\}$.


\section{Simulation}
\label{simulation}


We evaluate the frequency estimation performance of 2nd Taylor G-LASSO, compared with LASSO \cite{Tibshirani1996, Malioutov2005, Weiss2014}, Neighbor G-LASSO \cite{Abtahi2018}, and 1st Taylor G-LASSO \cite{Jagannath2013, Tan2014}. We adopt the root-mean squared error (RMSE) and the empirical probability of correct detection (PCD) as performance metrics, defined as in \cite{Steffens2018a}:
\begin{align*}
	\text{RMSE} = 10\log_{10}\left( \sqrt{\frac{1}{KQ}\sum_{k = 1}^{K}\sum_{q = 1}^{Q} \left( {\widehat{u}}_{k,q} - u_{k} \right)^{2} } \right)
\end{align*}
and ${\text{PCD}} = Q_{\text{suc}}/Q$, respectively, where ${\widehat{u}}_{k,q}$ denotes the frequency estimates of the $k$-th signal in the $q$-th Monte Carlo run, $Q$ is the total number of Monte Carlo trials, and $Q_{\text{suc}}$ is the number of trials where the frequency estimates $\left\{ \left. \widehat{u}_{k} \right| k = 1, 2, \cdots, K \right\}$ fulfill: $\max_{k} \left\{\left| \widehat{u}_{k} - {u}_{k} \right|\right\} \leq \delta/2$. The Cramér–Rao bound (CRB) \cite{Ma2021} is drawn as a benchmark for RMSE comparison.

\begin{figure}
\centerline{\includegraphics[width=\columnwidth]{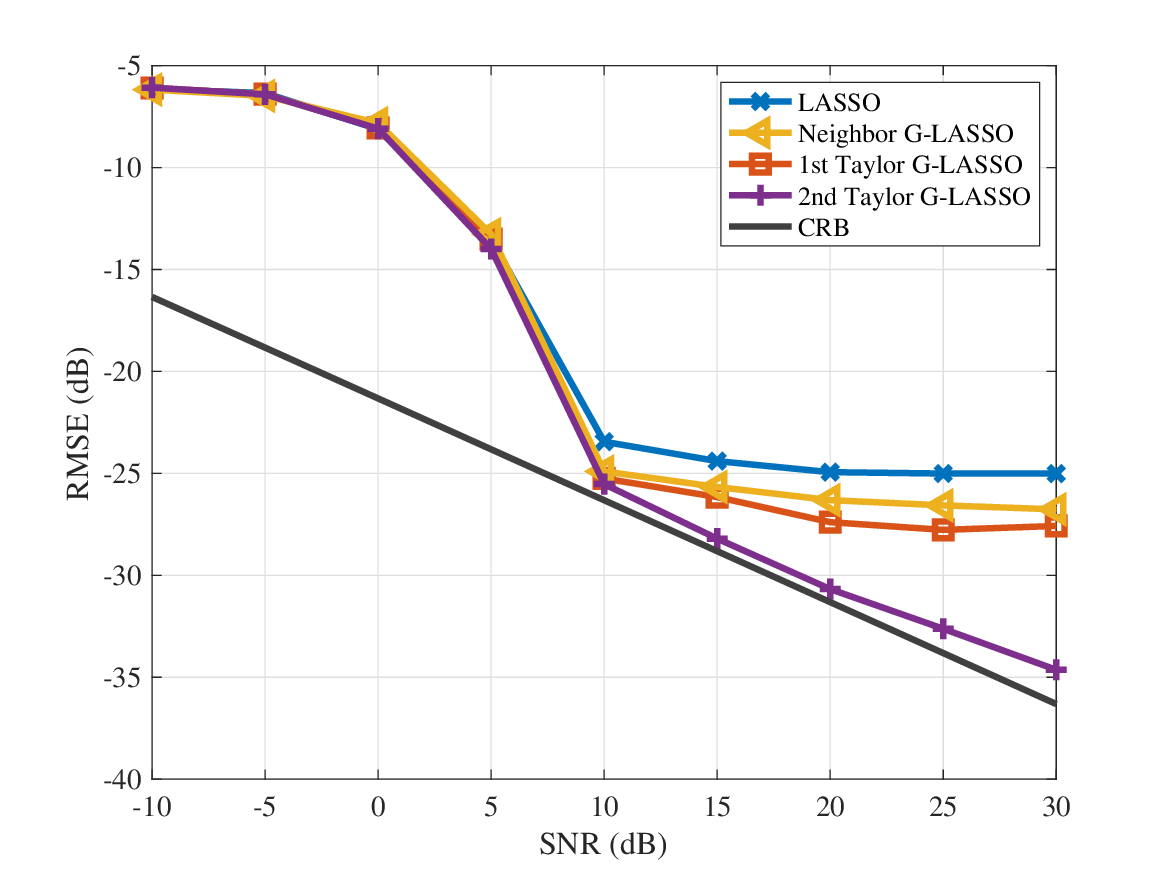}}
\caption{RMSE versus SNR with $M = 16$ sensors, $K = 2$ sources, $L = 200$ frequency grids, and grid size $\delta = 0.01$. }
\label{rmse_snr}
\end{figure}

\begin{figure}
\centerline{\includegraphics[width=\columnwidth]{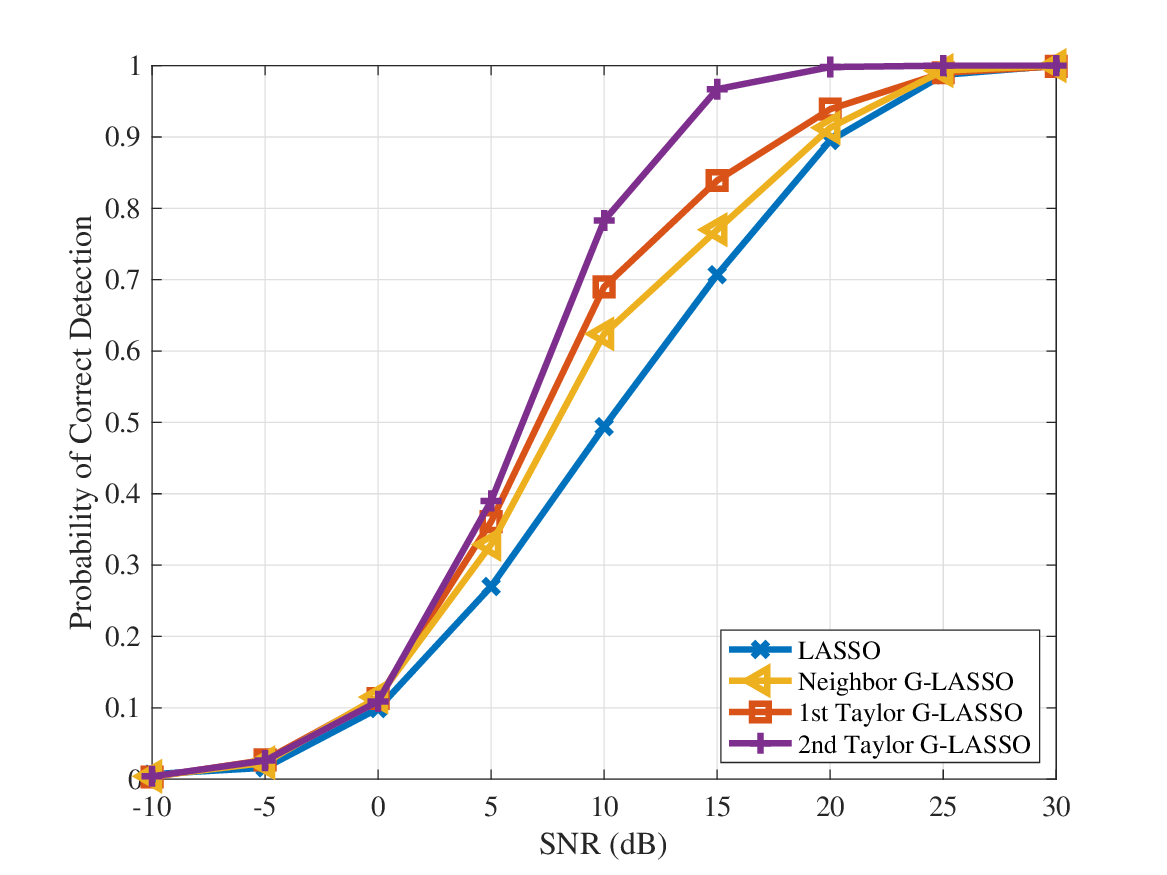}} 
\caption{PCD versus SNR with $M = 16$ sensors, $K = 2$ sources, $L = 200$ frequency grids, and grid size $\delta = 0.01$. }
\label{pcd_snr}
\end{figure}

In the first experiment, a linear array of $M = 16$ omnidirectional sensors is considered to receive $K = 2$ signals with spatial frequencies ${\bm u} = [0.1815 , 0.7942]^{\text{T}}$. {The $M = 16$ sensors are randomly selected from a ULA of $20$ sensors with half-wavelength inter-element spacing.} The frequency grid size is set to be $\delta = 0.01$, and hence the number of grids is $L = 200$. That is, the preset frequency grids are $\{-1, -0.99, \cdots, 0.98, 0.99\}$. Two parameters utilized in (\ref{proposedconvex}) are given as $\eta = 10^{-5}$ and $\mu = \sigma\sqrt{M\ln(M)}$ \cite{Bhaskar2013} with $\sigma$ denoting the standard deviation of the noise vector, which is assumed to be known \textit{a priori} in our simulations. $Q = 1000$ Monte Carlo trials are performed. The results of RMSE versus SNR and PCD versus SNR are plotted in Figures \ref{rmse_snr} and \ref{pcd_snr}, respectively. It is seen that, in the large SNR region, 2nd Taylor G-LASSO has significantly lower RMSE compared with the other grid-based approaches, and the PCD of 2nd Taylor G-LASSO is higher than those of the other tested methods.

%

\begin{figure}
\centerline{\includegraphics[width=\columnwidth]{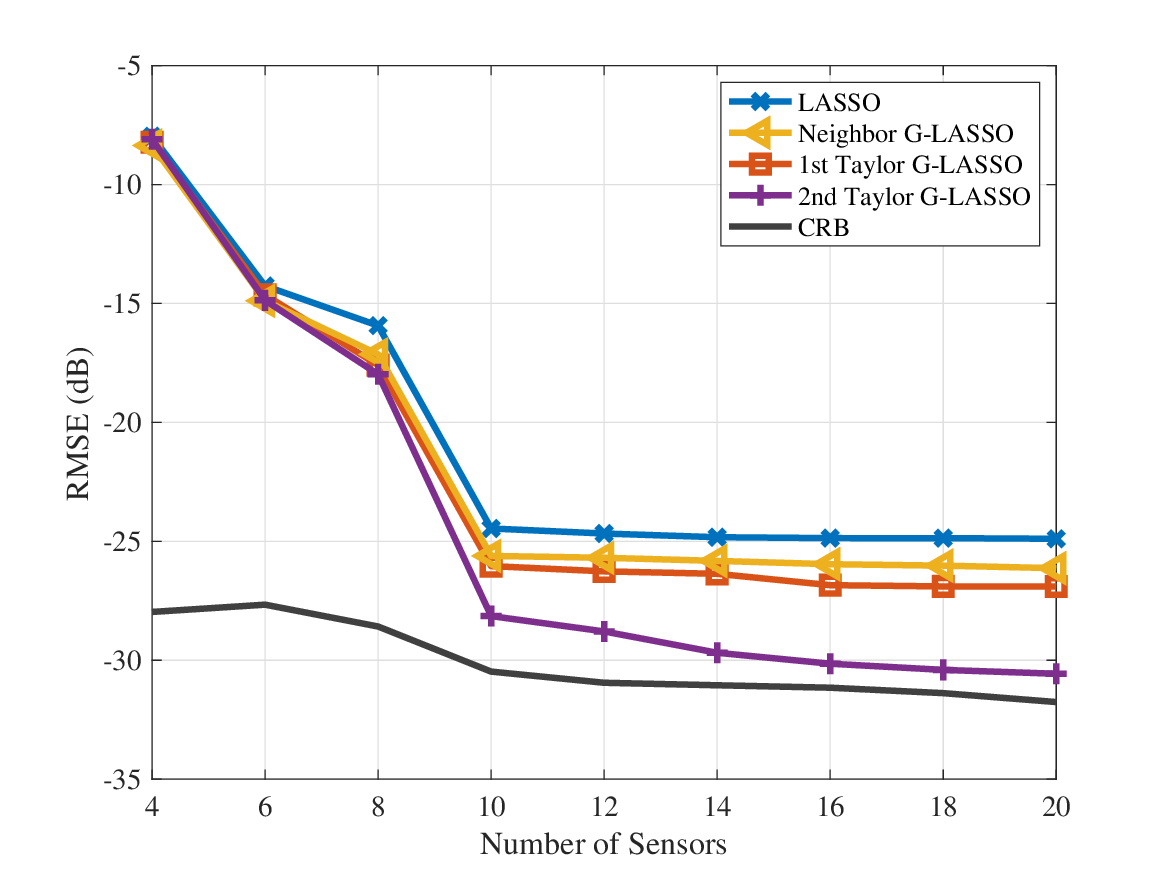}}
\caption{RMSE versus number of sensors with $\text{SNR} = 20 ~ \text{dB}$, $K = 2$ sources, $L = 200$ frequency grids, and grid size $\delta = 0.01$. }
\label{rmse_numbersensor}
\end{figure}

\begin{figure}
\centerline{\includegraphics[width=\columnwidth]{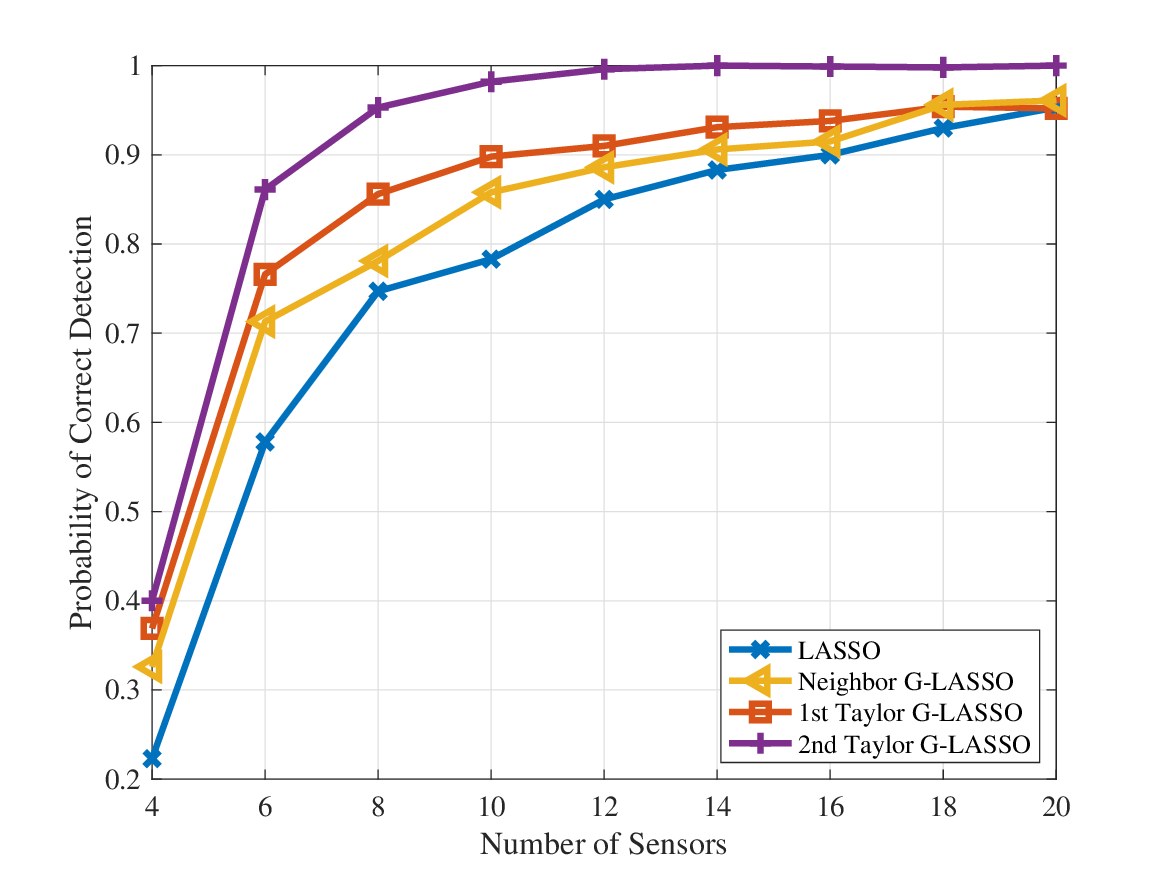}}  
\caption{PCD versus number of sensors with $\text{SNR} = 20 ~ \text{dB}$, $K = 2$ sources, $L = 200$ frequency grids, and grid size $\delta = 0.01$. }
\label{pcd_numbersensor}
\end{figure}

In the second experiment, {we randomly select $M$ sensors from a ULA of $20$ sensors with half-wavelength inter-element spacing, and $M$ varies from $4$ to $20$.} SNR is fixed to $20$ dB, while the remaining parameters are the same as those in the first experiment. The RMSE and PCD are depicted in Figures \ref{rmse_numbersensor} and \ref{pcd_numbersensor}, respectively. The results exhibit again better performance of the proposed 2nd Taylor G-LASSO than the other competitors.

%

\begin{figure}
\centerline{\includegraphics[width=\columnwidth]{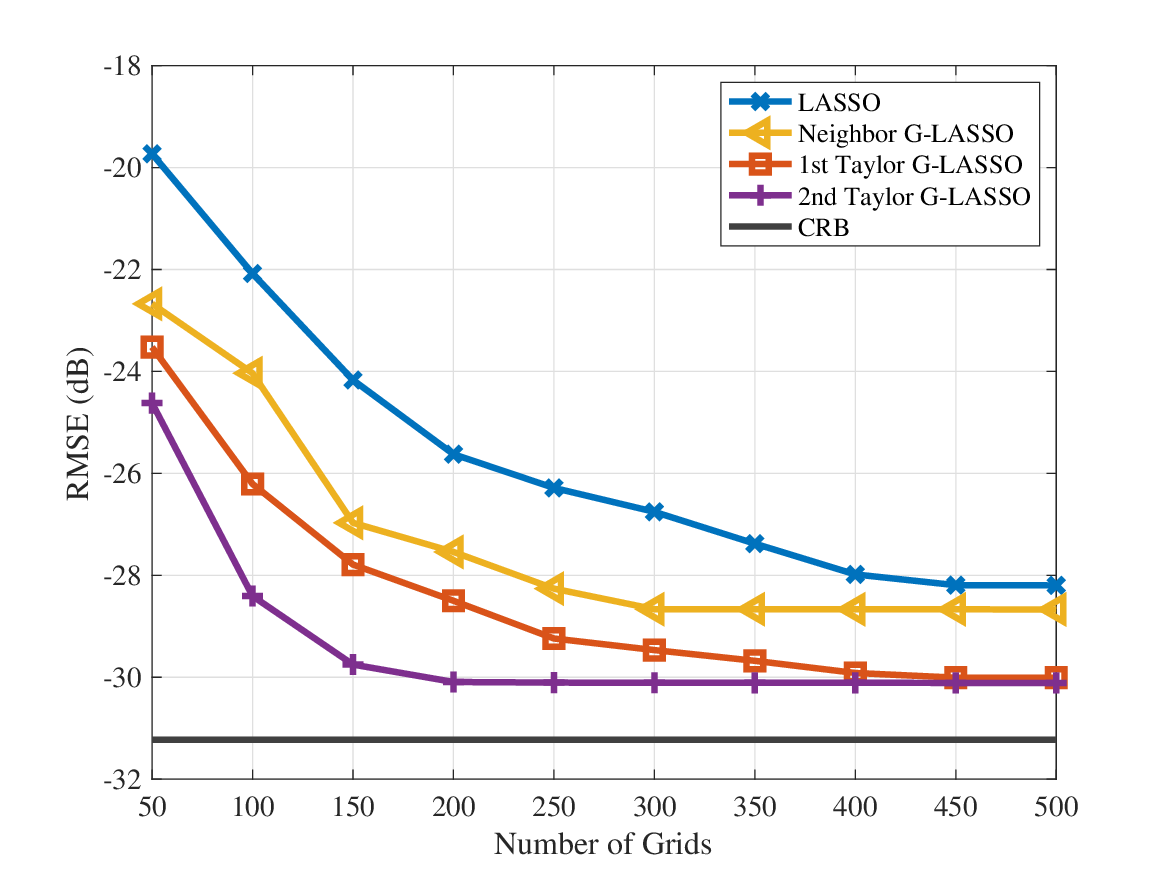}}
\caption{RMSE versus number of frequency grids with $\text{SNR} = 20 ~ \text{dB}$, $M = 16$ sensors, and $K = 2$ sources. }
\label{rmse_grid}
\end{figure}

\begin{figure}
\centerline{\includegraphics[width=\columnwidth]{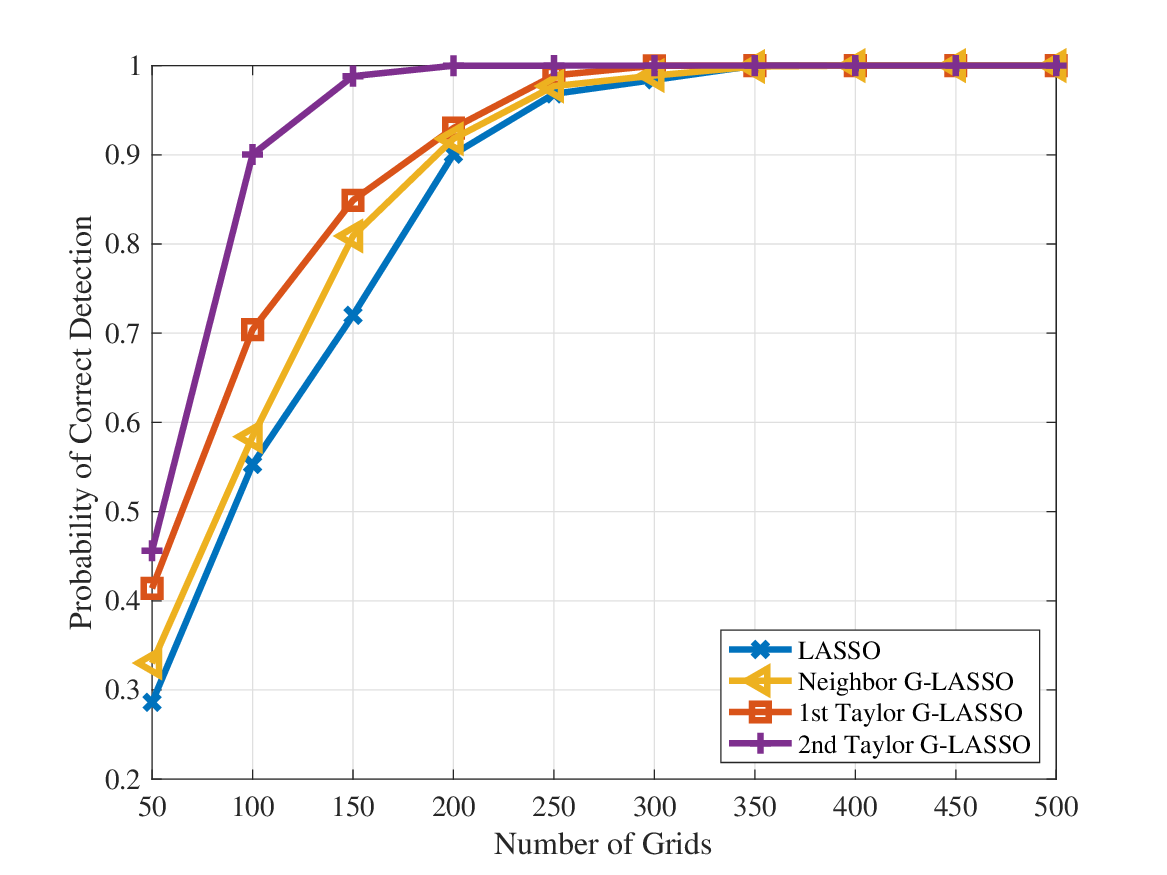}} 
\caption{PCD versus number of frequency grids with $\text{SNR} = 20 ~ \text{dB}$, $M = 16$ sensors, and $K = 2$ sources. }
\label{pcd_grid}
\end{figure}

{In the third experiment, the number of frequency grids, i.e., $L$, varies from $50$ to $500$ with a step size of $50$, the $\text{SNR}$ is fixed to $20 ~ \text{dB}$, while the other parameters are unchanged as those in the first experiment. The RMSE and PCD results are shown in Figures \ref{rmse_grid} and \ref{pcd_grid}, respectively. It can be seen that (i) When the number of grids is $L < 400$ (equivalently grid size of $\delta > 1/200$), the RMSE of 2nd Taylor G-LASSO is evidently smaller than those of the other tested methods; and (ii) When $L \geq 400$ (that is $\delta \leq 1/200$), the RMSE of 1st Taylor G-LASSO is very close to that of 2nd Taylor G-LASSO. This verifies that 2nd Taylor G-LASSO works better than 1st Taylor G-LASSO in terms of DOA estimation accuracy, especially when the grid size is not sufficiently small. }

%

\begin{figure}
\centerline{\includegraphics[width=\columnwidth]{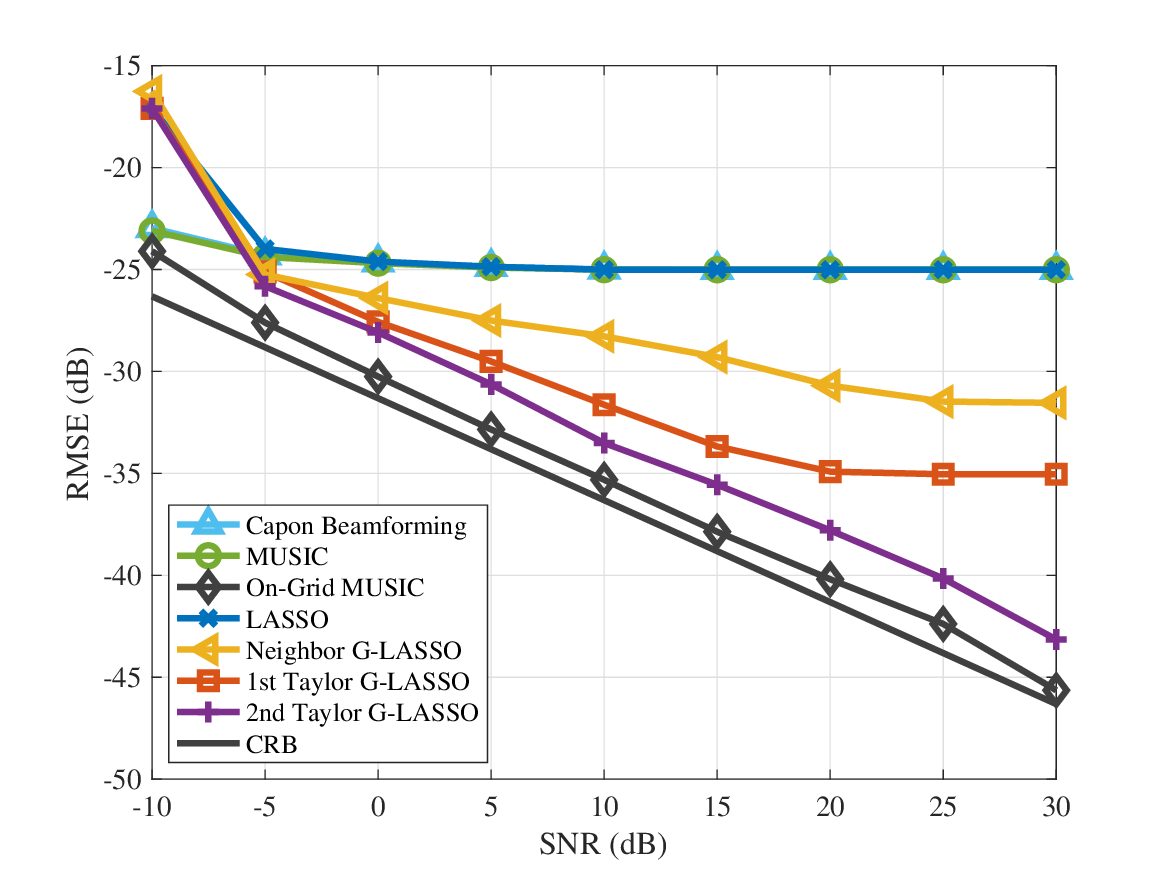}}
\caption{RMSE versus SNR in multiple-snapshot scenarios with $100$ snapshots, $M = 16$ sensors, and $K = 2$ sources. }
\label{rmse_snr_multiplesnapshot}
\end{figure}

\begin{figure}
\centerline{\includegraphics[width=\columnwidth]{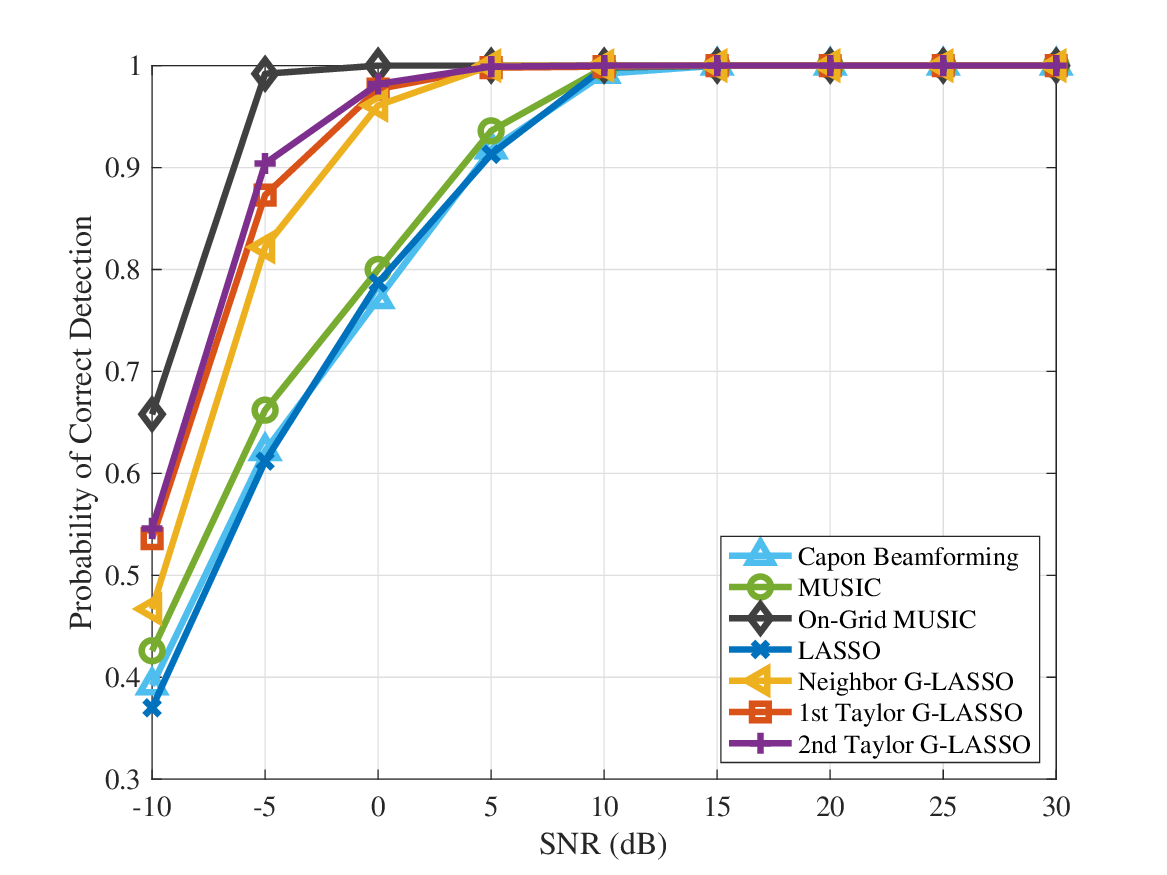}} 
\caption{PCD versus SNR in multiple-snapshot scenarios with $100$ snapshots, $M = 16$ sensors, and $K = 2$ sources. }
\label{pcd_snr_multiplesnapshot}
\end{figure}

{In the last experiment, we test the performances of the proposed method and several other algorithms in multiple-snapshot scenarios. We utilize $100$ snapshots, and the other parameters are set to be the same as those in the first experiment. The strategy of transforming the multiple-snapshot signal model into a single-snapshot one, which has been detailed in Remark \ref{multiple_to_single} in Section \ref{proposedminiprob}, is applied to LASSO, Neighbor G-LASSO, 1st Taylor G-LASSO, and 2nd Taylor G-LASSO. In addition, in this example, we also consider two classical methods, namely, the Capon beamforming and multiple signal classification (MUSIC) algorithms \cite{Chung2014}. For comparison, on-grid MUSIC with a much tinier grid size $\delta = 0.0001$ is also examined. The RMSE and PCD are plotted in Figures \ref{rmse_snr_multiplesnapshot} and \ref{pcd_snr_multiplesnapshot}, respectively, from which it is seen that both Capon beamforming and MUSIC algorithms share similar performance with LASSO in the off-grid setup. On-grid MUSIC has the smallest RMSE and the largest PCD among all the tested approaches. The proposed 2nd Taylor G-LASSO outperforms LASSO, Neighbor G-LASSO, and 1st Taylor G-LASSO. 
}

%

\section{Conclusion}

We have investigated the off-grid DOA estimation problem and have proposed a method using the second-order Taylor approximation. By exploring the properties of the block signal, we have added the proportionality relationship to our optimization problem. A Monte Carlo test has shown the usefulness of such proportionality relationship in the sense that it increases the probabilities of $\{\beta_{2K} < 1\}$ and $\{\beta_{2K} < \sqrt{2} - 1\}$. Numerical results have demonstrated that the proposed method outperforms several existing grid-based DOA estimation approaches.

\balance
\biboptions{numbers,sort&compress}
\bibliography{strings,refs}

\end{document}